\documentclass[
superscriptaddress,
preprint,
amsmath,amssymb,
]{revtex4-2}

\usepackage{graphicx}
\usepackage{dcolumn}
\usepackage{changepage} 
\usepackage{lipsum}

\usepackage{bm}
\usepackage{subfigure}
\usepackage{booktabs}
\usepackage{graphicx}
\usepackage{upgreek}
\usepackage{subfigure}
\usepackage{color}
\usepackage{amsmath}
\usepackage[percent]{overpic}
\usepackage{makecell}
\usepackage{lineno}
\begin{document}
\title{Electrically pumped  ultrabright entangled photons on chip}
\author{Xu-Feng Jiao}\thanks{These authors contributed equally to this work}
\affiliation{Hefei National Research Center for Physical Sciences at the Microscale and School of Physical Sciences,
University of Science and Technology of China, Hefei, Anhui 230026, China}
\affiliation{Jinan Institute of Quantum Technology and CAS Center for Excellence in Quantum Information and Quantum Physics, University of Science and Technology of China, Jinan, 250101, China}
\affiliation{Hefei National Laboratory, University of Science and Technology of China, Hefei, Anhui 230088, China}
\author{Ming-Yang Zheng}\thanks{These authors contributed equally to this work}
\affiliation{Jinan Institute of Quantum Technology and CAS Center for Excellence in Quantum Information and Quantum Physics, University of Science and Technology of China, Jinan, 250101, China}
\affiliation{Hefei National Laboratory, University of Science and Technology of China, Hefei, Anhui 230088, China}
\author{Yi-Hang Chen}
\affiliation{State Key Laboratory of Optoelectronic Materials and Devices, Institute of Semiconductors, Chinese Academy of Sciences, Beijing 100083, China.
}
\affiliation{Center of Materials Science and Optoelectronics Engineering, University of Chinese Academy of Sciences, Beijing 100049, China
}
\author{Bo Cao}
\affiliation{Jinan Institute of Quantum Technology and CAS Center for Excellence in Quantum Information and Quantum Physics, University of Science and Technology of China, Jinan, 250101, China}
\author{Xina Wang}
\affiliation{Hefei National Research Center for Physical Sciences at the Microscale and School of Physical Sciences,
University of Science and Technology of China, Hefei, Anhui 230026, China}
\affiliation{Jinan Institute of Quantum Technology and CAS Center for Excellence in Quantum Information and Quantum Physics, University of Science and Technology of China, Jinan, 250101, China}
\author{Yang Liu}
\affiliation{Jinan Institute of Quantum Technology and CAS Center for Excellence in Quantum Information and Quantum Physics, University of Science and Technology of China, Jinan, 250101, China}
\affiliation{Hefei National Laboratory, University of Science and Technology of China, Hefei, Anhui 230088, China}
\author{Cheng-Ao Yang}
\affiliation{State Key Laboratory of Optoelectronic Materials and Devices, Institute of Semiconductors, Chinese Academy of Sciences, Beijing 100083, China.
}
\affiliation{Center of Materials Science and Optoelectronics Engineering, University of Chinese Academy of Sciences, Beijing 100049, China
}
\author{Xiu-Ping Xie}
\affiliation{Jinan Institute of Quantum Technology and CAS Center for Excellence in Quantum Information and Quantum Physics, University of Science and Technology of China, Jinan, 250101, China}
\affiliation{Hefei National Laboratory, University of Science and Technology of China, Hefei, Anhui 230088, China}
\author{Chao-Yang Lu}
\affiliation{Hefei National Research Center for Physical Sciences at the Microscale and School of Physical Sciences,
University of Science and Technology of China, Hefei, Anhui 230026, China}
\affiliation{Hefei National Laboratory, University of Science and Technology of China, Hefei, Anhui 230088, China}
\author{Zhi-Chuan Niu}
\affiliation{State Key Laboratory of Optoelectronic Materials and Devices, Institute of Semiconductors, Chinese Academy of Sciences, Beijing 100083, China.
}
\affiliation{Center of Materials Science and Optoelectronics Engineering, University of Chinese Academy of Sciences, Beijing 100049, China
}
\author{Qiang Zhang}\altaffiliation{Contact author: qiangzh@ustc.edu.cn}
\affiliation{Hefei National Research Center for Physical Sciences at the Microscale and School of Physical Sciences,
University of Science and Technology of China, Hefei, Anhui 230026, China}
\affiliation{Jinan Institute of Quantum Technology and CAS Center for Excellence in Quantum Information and Quantum Physics, University of Science and Technology of China, Jinan, 250101, China}
\affiliation{Hefei National Laboratory, University of Science and Technology of China, Hefei, Anhui 230088, China}

\author{Jian-Wei Pan}
\affiliation{Hefei National Research Center for Physical Sciences at the Microscale and School of Physical Sciences,
University of Science and Technology of China, Hefei, Anhui 230026, China}
\affiliation{Hefei National Laboratory, University of Science and Technology of China, Hefei, Anhui 230088, China}
\date{\today}
\date{\today}


\maketitle
\begin{adjustwidth}{0.5cm}{0.5cm}

\textbf{Entangled photon sources (EPS) are essential for quantum science and technology. Despite advancements in integrated optical platforms like thin-film lithium niobate, a scalable, high-performance, chip-scale EPS has remained elusive. We address this by demonstrating an electrically pumped, post-selection-free polarization-EPS, achieved through hybrid integration of a distributed feedback laser with thin-film lithium niobate chip which integrates periodically poled lithium niobate waveguides, beam splitter, and polarization rotator combiner. By injecting current into the chip, we realize a high-performance EPS with a bandwidth of 73 nm and an entanglement pair generation rate of 4.5\(\times\)10\(^{10}\) pairs/s/mW. The polarization entanglement shows Bell-state fidelities above 96\% across frequency-correlated modes. This compact, integrated EPS enables key applications, including high-speed quantum key distribution via wavelength division multiplexing, satellite-based quantum communication, and entanglement-based quantum metrology.}
\end{adjustwidth}
\vspace{2em}

Quantum entangled photon sources play a pivotal role in both fundamental studies of quantum mechanics~\cite{nobel2022physics} and the advancement of quantum technologies, including quantum computation ~\cite{2001A,2005Experimental,doi:10.1126/science.abe8770,Alexander2025,aghaee2025scaling},
quantum communication ~\cite{PhysRevLett.67.661,PhysRevLett.69.2881,doi:10.1126/science.aan3211,2018An,zhuang2024ultrabright},  quantum metrology ~\cite{Dowling01032008,kalashnikov2016infrared,doi:10.1126/science.adk7825,defienne2024advances}.
In recent years, significant breakthroughs has been made in these fields. Entangled photon sources have been leveraged to develop quantum processors that have demonstrated quantum supremacy ~\cite{doi:10.1126/science.abe8770,Alexander2025,aghaee2025scaling}, satellite based 1000 km quantum key distribution ~\cite{doi:10.1126/science.aan3211}, and long distance fiber based quantum network ~\cite{2018An,zhuang2024ultrabright}.  
Concurrently, the limitations in Mid-infrared spectroscopy measurement ~\cite{kalashnikov2016infrared} 
and high quality imaging ~\cite{doi:10.1126/science.adk7825} have also been overcame using quantum entanglement.

Over the past few decades, the main way to generate entangled photons are based on the nonlinear optic process in the bulk crystal and waveguide \cite{PhysRevLett.75.4337,shih2003entangled,PhysRevLett.100.183601,PhysRevLett.104.253602,ref35,yao2012observation,ref31,PhysRevLett.120.140405,zhuang2024ultrabright,PhysRevLett.121.200502}.  
In recent years, the development of integrated EPS has received widespread attention through photonic-on-insulator (PIC) architectures, enabling small footprint
and high stability beyond traditional tabletop setups ~\cite{silverstone2014chip,Atzeni:18,ref21,PhysRevLett.124.163603,wang2021integrated,wang2021integrated,ref22,ref24,ref20,jia2025continuous}.
However, those schemes are realized with optical pumping in which a bulky external laser is inevitably required, resulting in large system size and poor scalability.
To achieve an on-chip entanglement source, one promising solution is to adopt the electrically pumped approach based on the integration of pump laser and PIC chip. Electrically pumped time-bin entangled photon pair has been achieved via hybrid integration of InP and Si\(_{\mathrm{3}}\)N\(_{\mathrm{4}}\) based on spontaneous four-wave mixing (SFWM) ~\cite{mahmudlu2023fully}, however, challenges remain in terms of brightness and bandwidth performance, making it unable to meet the requirements of widespread applications. Meanwhile, it is in general harder to achieve integrated polarization entanglement source comparing to time bin source.

Among the PIC platforms, including silicon ~\cite{ref21,silverstone2014chip,wang2021integrated}, Si\(_{\mathrm{3}}\)N\(_{\mathrm{4}}\) ~\cite{ref20,wang2021integrated,mahmudlu2023fully,jia2025continuous}, and lithium niobate ~\cite{PhysRevLett.124.163603,wang2021integrated,ref22,ref24},the thin-film lithium niobate (TFLN) platform has emerged as the most promising approach for realizing the target EPS featuring high-performance and enhanced scalability. This is primarily due to the exceptional properties of lithium niobate, such as high second-order nonlinearity, broad transparency window, flexible domain and dispersion engineering, which are beneficial for efficient spontaneous parametric down-conversion (SPDC) process. Recent advancements leveraging TFLN platform have demonstrated the potential to construct compatible EPS ~\cite{PhysRevLett.124.163603,ref22,ref24},
while achieving an on-chip EPS that simultaneously exhibits high integration and high performance remains a challenge.

\begin{figure*}[!htbp]
\centering
\includegraphics[width=0.9\linewidth]{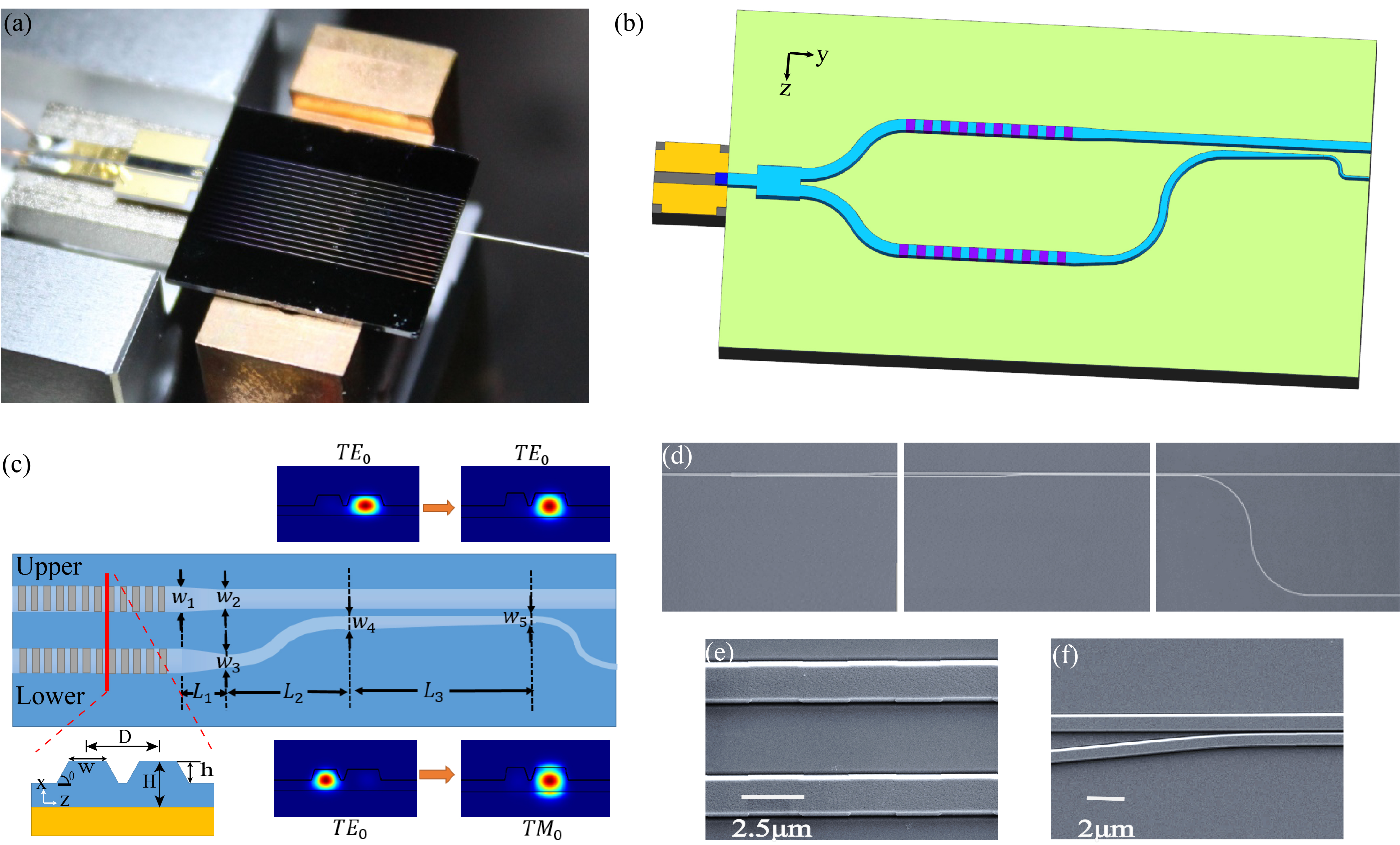}
\caption{Characterization of the electrically pumped polarization-EPS: (a) photo of the hybrid integrated EPS with DFB and TFLN chip. Lensed fiber is used for collecting output lights. (b) Schematic illustration of the polarization-EPS by hybrid integration of DFB and TFLN chips. (c) Scanning electron microscope (SEM) image of the dual channel PPLN waveguides. (d) SEM image of the PRC.  (e) Simulated GVD in PPLN  waveguides and the cross section of the dual channel PPLN waveguides with parameters of W = 1.41 \(\mu\)m, D = 5 \(\mu\)m, H = 0.6 \(\mu\)m, h= 0.3 \(\mu\)m. (f) Schematic structure of the  PRC and the optical mode profiles at different positions of chip. \(w_{1}=1.41\) \(\mu\)m, \(w_{2}=1.4\) \(\mu\)m, \(w_{3}=1\) \(\mu\)m, \(w_{4}=1\) 
 \(\mu\)m, \(w_{5}=0.8\)  \(\mu\)m, \(L_{1}=250\) \(\mu\)m, \(L_{2}=50\)  \(\mu\)m, \(L_{3}=6000\)  \(\mu\)m. }
\label{fig:structure}
\end{figure*}

In this work, we present an electrically pumped, post-selection-free hybrid-integrated polarization-EPS by combining a DFB laser with a TFLN chip. The device achieves a pair generation rate (PGR) of 4.5\(\times\)10\(^{10}\) pairs/s/mW and a bandwidth of 73 nm, which is six orders  higher and one order higher than the previous result, respectively~\cite{mahmudlu2023fully}. The broad bandwidth enables multi-channel polarization-entangled photon generation, which is a critical feature for high-dimensional quantum information protocols. Additionally, we have demonstrated a high fidelity surpassing 96\% across different frequency modes.

\textit{Design and fabrication}—To construct the electrically pumped polarization-EPS depicted in Fig.~\ref{fig:structure}(a), we integrate a single-mode DFB laser chip with a TFLN photonic circuit using butt-coupling techniques~\cite{ref27,ref26,ref28}.  The hybrid integration scheme, illustrated schematically in Fig.~\ref{fig:structure}(b), ensures efficient optical mode transfer between the two components. The DFB laser emits horizontally polarized pump light at \(\sim\) 780 nm, achieving a side-mode suppression ratio exceeding 40 dB and a polarization extinction ratio \(>\) 20 dB. The emission wavelength exhibits a linear temperature-dependent shift of 0.067 nm/°C, which enables precise wavelength tuning through thermal adjustment.

The TFLN photonic circuit is designed and fabricated on an x-cut TFLN platform, which consists of a 600 nm thick lithium niobate layer and a 2 \(\mu\)m thick buried silicon oxide layer on a silicon substrate. The TFLN chip integrates a multi-mode interference (MMI) beam splitter, dual-channel periodically poled lithium niobate (PPLN) waveguides, and a polarization rotator combiner (PRC) ~\cite{ref38,ref30,ref29}, realizing parametric photon generation and interference. A symmetrically designed 1\(\times\)2 MMI splitter with an simulated insertion loss of  3.1 dB ensures equal division of the coupled pump light into both channels. 
This configuration ensures both PPLN waveguides are pumped by the same laser source. Two-channel pump beams are injected into two identical 6-mm-long PPLN waveguides with first order type-0 quasi phase matching periods of 4.13 \(\mu\)m. The dual-waveguide have a center-to-center distance of 5 \(\mu\)m and width of 1.41 \(\mu\)m.
Photon pairs generated in fundamental (T\(\rm{E_0}\)) modes are combined and polarization-converted by the on-chip PRC.
A 300 \(\mu\)m-long transition region is incorporated prior to the PRC to facilitate the connection with the PPLN waveguides. The upper waveguide linearly decreases in width from 1.41 \(\mu\)m to 1.4 \(\mu\)m over a 250 \(\mu\)m propagation length, followed by a 50 \(\mu\)m straight segment maintaining a constant width of 1.4 \(\mu\)m. The lower waveguide similarly tapers from 1.41 \(\mu\)m to 1 \(\mu\)m within 250 \(\mu\)m, followed by an S-bend structure for a subsequent 50 \(\mu\)m section with a constant width of 1 \(\mu\)m.
In the 6 mm-long PRC region, the upper and lower waveguides are separated by an edge-to-edge distance of 0.5 \(\mu\)m. 
The horizontal polarization of the photon pairs \(\mathrm{|H_{s}H_{i}\rangle}\) maintains in the upper waveguide with a fixed width of 1.4 \(\mu\)m, while the horizontally polarized photon pairs in the lower waveguide \(\mathrm{|H_{s}H_{i}\rangle}\) is converted to vertically polarized photon pairs \(\mathrm{|V_{s}V_{i}\rangle}\) through width tapering from 1.0 \(\mu\)m to 0.8 \(\mu\)m, and coupled to the upper waveguide as shown in  Fig.~\ref{fig:structure} (c).
Here the horizontally polarized state \(\mathrm{|H\rangle}\) and vertically polarized state \(\mathrm{|V\rangle}\) correspond to the T\(\rm{E_0}\) mode and transverse magnetic fundamental (T\(\rm{M_0}\)) mode, respectively. 
The PRC yields an output polarization state characterized by:
\begin{equation}
\left|\psi\right\rangle = \alpha\left|\mathrm{H_s H_i}\right\rangle + \beta e^{i\phi}\left|\mathrm{V_s V_i}\right\rangle,
\label{eq:PES}
\end{equation}
where \(\alpha\) and \(\beta\) denote the amplitudes of the \(\mathrm{|H_{s} H_{i}\rangle}\) and \(\mathrm{|V_{s} V_{i}\rangle}\) photon states, and phase difference \(\phi=\phi_{\mathrm{H}}-\phi_{\mathrm{V}}\) is determined by the optical path difference between photons originating from the upper and lower waveguide channels. Achieving the maximally entangled state (1/\({\sqrt{2}} \))(\(\mathrm{|H_{s} H_{i}\rangle}\) \(\pm\) \(\mathrm{|V_{s} V_{i}\rangle}\)) requires a 50: 50 splitting ratio in the fabricated MMI structure, identical SPDC processes in dual-PPLN waveguides, and balanced loss characteristics in the PRC structure across both waveguides.

Following the design, we have successfully implemented chip fabrication utilizing electron beam lithography (EBL)-based semiconductor processing technology. To fully characterize the performance of the on-chip integrated entanglement source, the independent components including MMI, PPLN waveguides, and PRC are individually characterized (see Supplemental Material for details of chip fabrication and component measurements).

\textit{Ultrabright polarization-entangled photons}—To characterize the generated polarization-entangled photon pair, the TFLN waveguide is pumped by a 779.5 nm DFB laser with 0.15 mW coupled power. The SPDC photons with wavelengths shorter (longer) than the degenerate wavelength of 1559 nm are designated as signal (idler) photons. A polarizer is employed to differentiate photons with different polarization emitted from the two waveguides. After passing through the polarizer, the photons are directed into an InGaAs CCD spectrometer via a multi-mode fiber and the absence of frequency components beyond 1600 nm is due to the reduced quantum efficiency of the InGaAs CCD spectrometer at longer wavelengths.
Figure 2 depicts the measured TE-TM mode conversion efficiency of the PRC (red dashed curve) alongside the SPDC photon pair spectrum (blue solid curve). Achieving broadband EPS necessitates concurrent broadband operation in both SPDC generation and PRC conversion processes (see supplementary). By engineering the initial and terminal widths of the PRC waveguide, we obtained a 160 nm operational bandwidth. 
As a result, the idler photons have an 80 nm bandwidth, while the signal photons have a 73 nm bandwidth.
\begin{figure}[ht]
\centering
\includegraphics[width=0.7\linewidth]{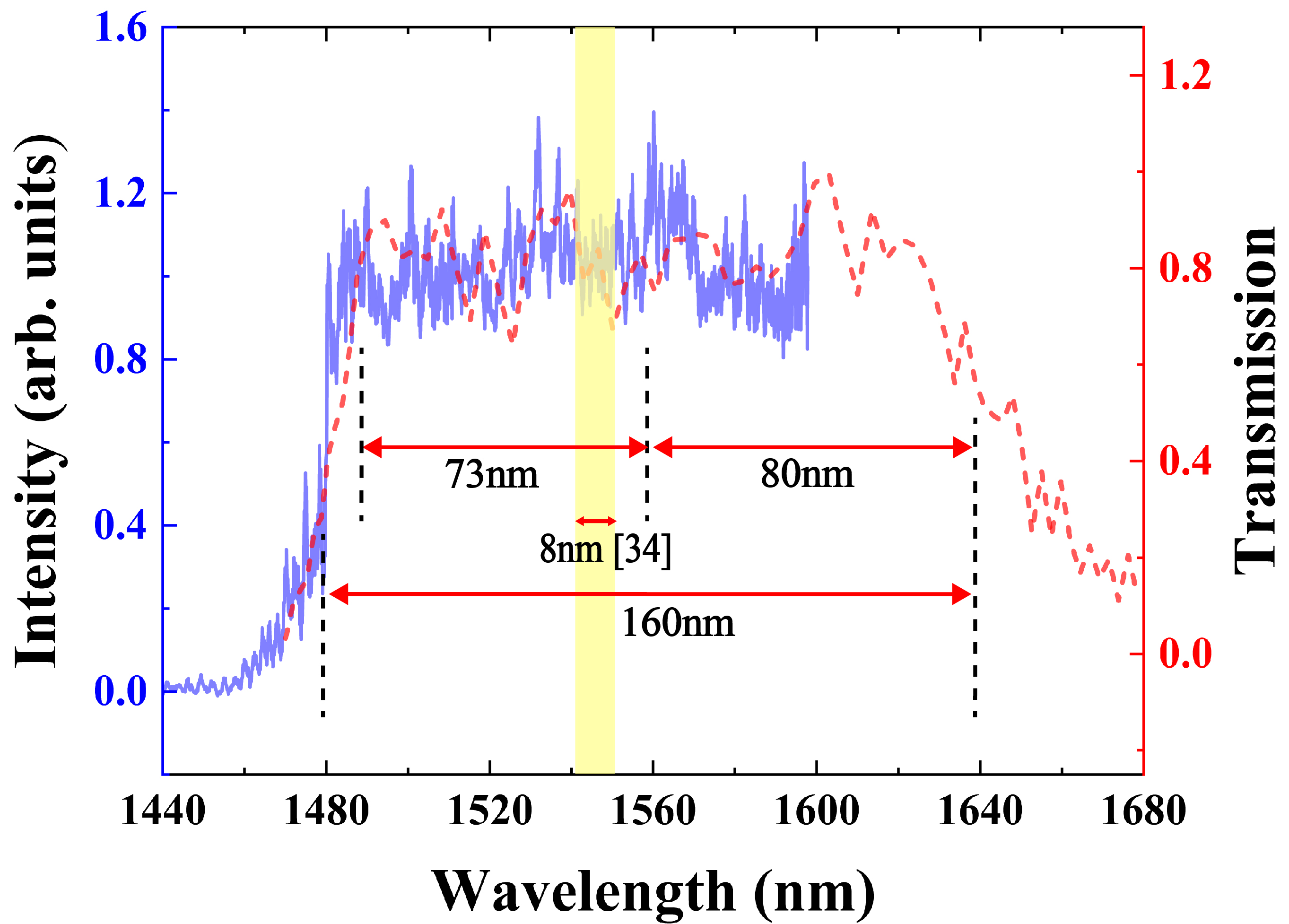}
\caption{Measured spectrum for vertically polarized photons (blue, solid) and measured transmission curves for TE-TM mode conversion and combination (red, dashed). The highlighted region shows the bandwidth of the electrically pumped entangled source demonstrated in Ref. [34].
}
\label{fig:prc+spectrum}
\end{figure}

\begin{figure*}[htp]
\centering
\includegraphics[width=1\linewidth]{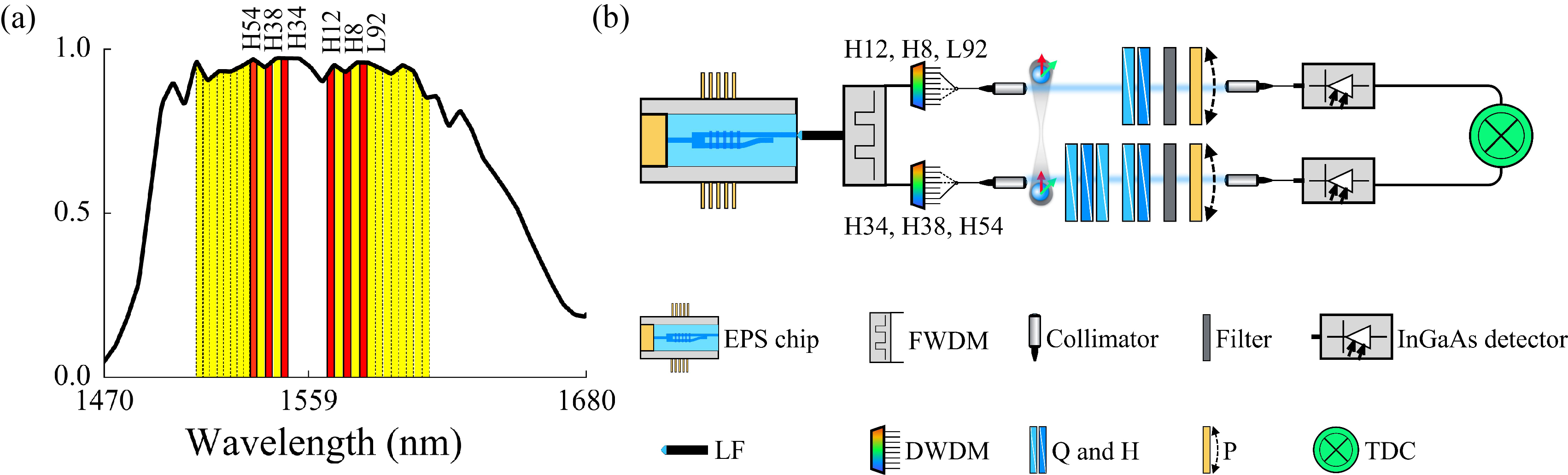}
\caption{
(a) Polarization-EPS spectrum with three sets of selected frequency channels. (b) The schematic diagram of experimental setup. 
    EPS, entangled photon source;
    TEC, thermoelectric cooler;
    LF, lense fiber with 2.5 \(\mu\)m mode field diameter (MFD);
    FWDM, filter Wavelength Division Multiplexing;
    DWDM, dense Wavelength Division Multiplexer;
    Filter, 850 nm long pass filter;
    Q, quarter-Wave Plate;
    H, half-Wave Plate;
    P, polarizer. 
}\label{fig:QST_setup}
\end{figure*}

Spectral separation of the frequency-correlated photon pairs emitted from the TFLN chip is required to characterize the brightness of the polarization-EPS.
A filter wavelength-division multiplexer (FWDM) is employed in conjunction with two dense wavelength-division multiplexer (DWDM) modules shown in Fig. 3(b), each featuring a 200 GHz transmittance bandwidth. These components facilitated spatial discrimination between idler and signal photons. Adhering to the principle of energy conservation, the selected DWDM channels for the idler and signal photons feature transmission wavelengths of H8/H38 (1571.24 nm and 1546.92 nm), respectively. After the DWDM, photons are coupled into single mode fibers and subsequently directed into InGaAs single photon detectors. The single count events and coincidence events are recorded using a time to digital converter (IDQ, ID1000) with a coincidence window of 2 ns. With an on-chip pump power of 18.8 \(\mu\)W,
the recorded coincidence counting rate is \(R_{\mathrm{CC}}\) = 1.1 kHz, whereas the accidental coincidence rate is \(R_{\mathrm{AC}}\) = 29 Hz. The corresponding single counting rates for signal and idler photons are \(R_{\mathrm{s}}\) = 215 kHz and \(R_{\mathrm{i}}\) = 92 kHz,
respectively. Brightness of the source is estimated by the formula: \(B\) = \(R_{\mathrm{s}}R_{\mathrm{i}}\) / ((\(R_{\mathrm{AC}}-R_{\mathrm{CC}}) \cdot P \cdot \Delta\lambda\)),
where \(P\) is on-chip pump power and \(\Delta\lambda\) is the bandwidth of the DWDM\@.
The estimated brightness is 6.2 \(\times\) 10\(^8\) pairs/s/nm/mW, resulting in a PGR of \(4.5 \times 10^{10}\) pairs/s/mW calculated with a bandwidth of 73 nm. 
Compared with the first electrically pumped on-chip entangled sources based on Si\(_{\mathrm{3}}\)N\(_{\mathrm{4}}\) platforms utilized micro-resonator via SFWM ~\cite{mahmudlu2023fully}, our on-chip EPS has achieved a breakthrough with bandwidth increasing by nearly an order of magnitude and brightness by six orders of magnitude. 

\begin{figure}[ht]
\centering
\includegraphics[width=0.6\linewidth]{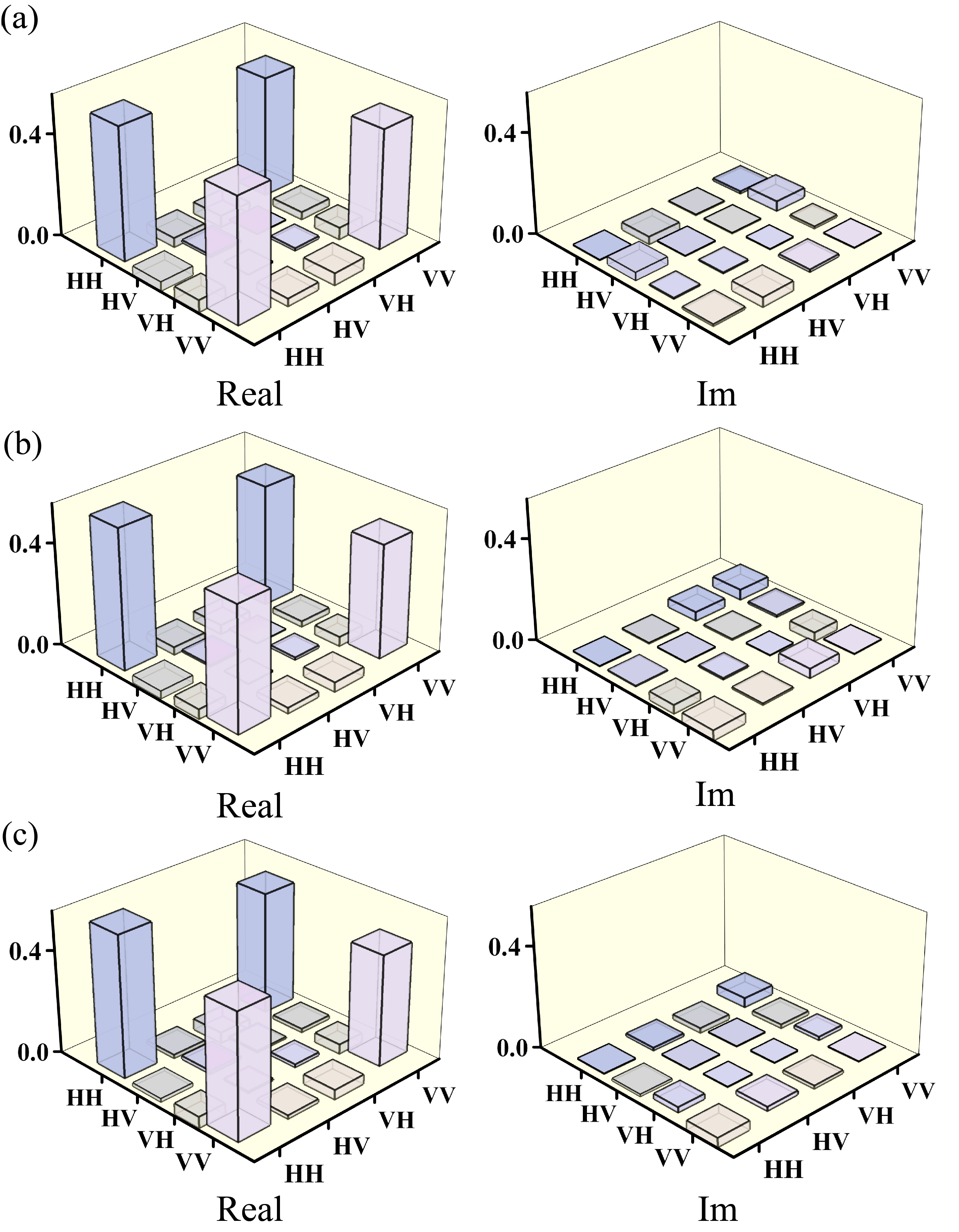}
\caption{
    Density matrixes of measured polarization entangled states.
    (a) In the channels L92/H54 (1584.53 nm and 1534.25 nm), the fidelity \(F\) = 0.965.
    (b) In the channels H8/H38 (1571.24 nm and 1546.92 nm), \(F\) = 0.971.
    (c) In the channels H12/H34 (1567.95 nm and 1550.12 nm), \(F\) = 0.969.
}\label{fig:QST_results}
\end{figure}

\textit{Polarization entanglement characterization}—To characterize the polarization entangled state generated by our hybrid integrated chip, we carry out a polarization correlation measurement using the experimental setup shown in Fig.~\ref{fig:QST_setup}(b). The DFB laser is coupled into the TFLN chip and a single-mode lensed fiber is used to collect the photons at the output port.
To demonstrate our broadband entanglement source’s potential for multi-frequency mode multiplexing, we quantitatively analyzed three selected
channels, as shown in Fig. 3(a). Specifically, the wavelength configurations are L92 and H54 (1584.53 nm and 1534.25 nm), H8 and H38 (1571.24 nm and 1546.92 nm), H34 and H12 (1567.95 nm and 1550.12 nm).
To compensate for the phase difference induced by the path length disparity between the lower and upper waveguides on the TFLN chip, a set of optics consisting of two QWPs and one HWP is implemented. For polarization correlation detection between signal and idler photons in the horizontal, vertical, and \(\pm\) 45\(^{\circ}\) linear bases, two polarizers rotate the light to align with these linear polarization directions. The observed polarization correlation is close to that expected for the ideal Bell state. The measured visibilities for each set of frequency channels are:  V\(_\mathrm{H-V}\) \(=\) 0.982 and V\(_{\mathrm{A-D}}\) \(=\) 0.975 in the channels H8/H38 (1571.24 nm and 1546.92 nm), V\(_\mathrm{H-V}\) \(=\) 0.978 and V\(_{\mathrm{A-D}}\) \(=\) 0.972 in the channels H12/H34 (1567.95 nm and 1550.12 nm),  V\(_\mathrm{H-V}\) \(=\) 0.976 and V\(_{\mathrm{A-D}}\) \(=\) 0.971 in the channels L92/H54 (1584.53 nm and 1534.25 nm) after subtracting
 the accidental coincidences, suggesting the violation of the CHSH inequality (V \(>\) 0.71).
\smallskip

In addition, quantum state tomography measurements are performed to fully evaluate the quantum state properties~\cite{james2001measurement}. 
By recording the coincidence counts for all 16 possible polarization basis combinations across all wavelength configurations, we reconstruct the density matrix \(\hat{\rho}\) of the quantum state using maximum likelihood estimation. The real parts and the imaginary parts of the density matrix for all wavelength channels are illustrated in Fig.~\ref{fig:QST_results}. The diagonal and off-diagonal elements between \(|\mathrm{H}\mathrm{H}\rangle\) and
\(|\mathrm{V}\mathrm{V}\rangle\) states dominate the density matrix, indicating that the state is close the Bell state.
The fidelity of our experimentally obtained state to an ideal Bell state can by estimated as: 
\begin{equation}
    F = \langle \Psi^{+} | \hat{\rho} | \Psi^{+} \rangle,
\end{equation}
where \(\Psi^{+} = (1/\sqrt{2}) (|\mathrm{HH}\rangle + |\mathrm{VV}\rangle)\) is an ideal Bell state.
For each pair of wavelength channels, the results are \(F\) = 0.965 (L92/H54), 0.971 (H38/H8), and  0.969 (H34/H12) after subtracting the accidental coincidences, which are very close to an ideal Bell state.

\textit{Conclusion and discussion}—
By integrating DFB laser with TFLN chip, we have successfully developed a scalable polarization-entangled  source in a compact hybrid platform (15 \(\times\) 20 $\rm{mm}^2$).  
Our integrated on-chip solution has demonstrated exceptional performance metrics without low-temperature refrigeration~\cite{Zhang2015,Liu2025}, including high PGR of 4.5\(\times\)10\(^{10}\) pairs/s/mW, broad bandwidth spanning 73 nm and high fidelity exceeding 96\% across all selected frequency channels, confirming the generation of high-quality entangled photon pairs.
Notably, the generation of such entangled sources could be powered by a portable battery, which significantly enhances its practicality and applicability across various fields like high-speed quantum key distribution via wavelength division multiplexing, satellite-based quantum communication, and entanglement based quantum metrology.

To advance multi-channel, high-density on-chip entangled photon pair generation, future efforts should focus on integrating on-chip pump laser via wafer bonding, micro-transfer printing, or potentially molecular beam epitaxy techniques. Furthermore, fully integration of quantum source generation, manipulation, networking, and detection on a TFLN platform would enable chip-scale quantum systems, supporting more applications in quantum science and technology.

\textit{Acknowledgments}—We thank Jun-Qiu Liu, Yang-Fan Jiang and Yuan Cao for insightful discussions. This work has been supported by the Innovation Program for Quantum Science and Technology (Grant No. 2021ZD0300800), the National Natural Science Foundation of China (Grant No. T2125010), the Key R\&D Plan of Shandong Province (Grant No. 2021ZDPT01), Natural Science Foundation of Shandong Province (Grant No. ZR2021LLZ013, No. ZR2022LLZ009, No. ZR2022LLZ010), the National Science Foundation of China (Grant No. 62204238), the Chinese Academy of Sciences Project for Young Scientists in Basic Research (Grant No. YSBR-112), the Beijing Nova Program (Grant No. 20240484621), the ``Announce the list and take charge" of the Major Special Plan of Science and Technology in Shanxi Province (Grant No. 202201030201009). Q. Z. were supported by the New Cornerstone Science Foundation through the Xplorer Prize, the SAICT Experts Program, the Taishan Scholar Program of Shandong Province, and Quancheng Industrial Experts Program. M.-Y. Z. were supported by the Taishan Scholar Program of Shandong Province and Haiyou Plan Project of Jinan.  

\bibliography{ref}

\newpage
\thispagestyle{empty}
\begin{center}
\vspace*{2cm}
{\LARGE\textbf{Supplementary Materials for: Electrically pumped ultrabright entangled photons on chip
}}\\[1cm]
\end{center}
\vspace{2cm}

\section*{Supplementary Note 1: Fabrication process}

The fabrication process of the TFLN chip comprises two major steps: periodic  poling and waveguide patterning. Initially, aluminum comb-shaped electrodes are fabricated via UV-based photolithography followed by wet chemical etching. An external electric field is then applied to induce periodic domain inversion within the crystal lattice. The resulting periodic polling structure can be observed in the SEM image shown in Fig.1(c). In this characterization, a second TFLN chip from the same fabrication batch after treatment with hydrofluoric acid is used. Subsequently, the waveguide patterns are fabricated via electron beam lithography (EBL) utilizing FOX16 photoresist, followed by inductively coupled plasma (ICP) etching to transfer these patterns into the lithium niobate layer. This process achieves an etch depth of \(\sim\) 300 nm and sidewall angles of \(\sim\) 63°.

\section*{Supplementary Note 2: MMI measurement}

\begin{figure}[htp]
\centering
\includegraphics[width=0.6\linewidth]{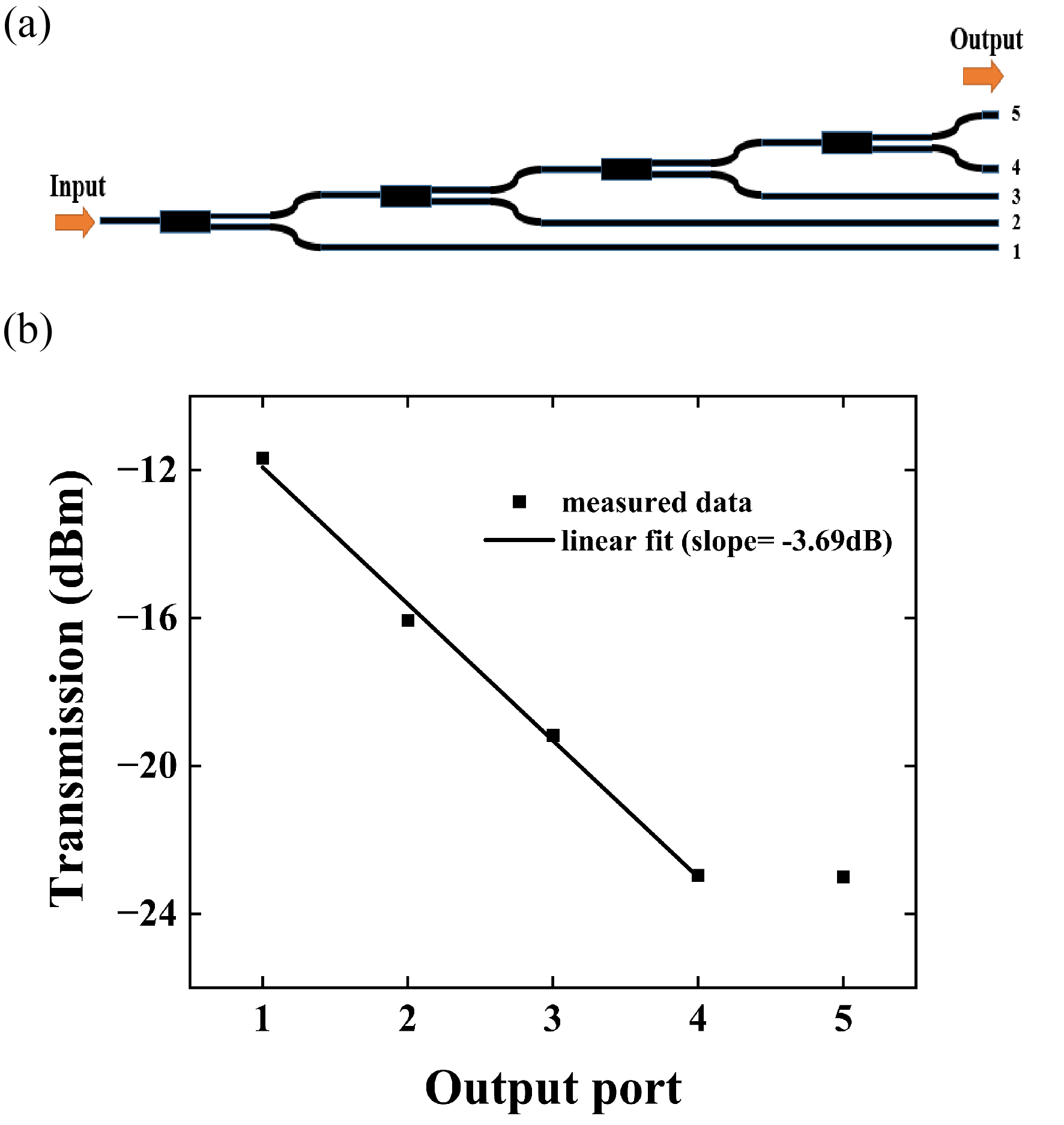}
\caption{
    (a) Schematic of MMI-tree structure. (b) Measurement results for different MMI-tree outputs.
}\label{fig:MMI}
\end{figure}
To investigate the insertion loss characteristics of an MMI for 780 nm, we employ an MMI-tree structure with a total length of 6 mm as illustrated in Fig.~\ref{fig:MMI}(a). The configuration consists of five output waveguides, each featuring uniform waveguide lengths but incorporating distinct numbers of MMI splitters. Experimental characterization reveals that the MMI device exhibited an insertion loss of  \(\sim\)3.47 dB with a splitting ratio less than less than 49\(\%\) : 51\(\%\) as shown in Fig.~\ref{fig:MMI}(b), indicating that amplitude distortion induced by the MMI can be effectively neglected.

\section*{Supplementary Note 3: SHG measurement}

\begin{figure}[htp]
\centering
\includegraphics[width=0.9\linewidth]{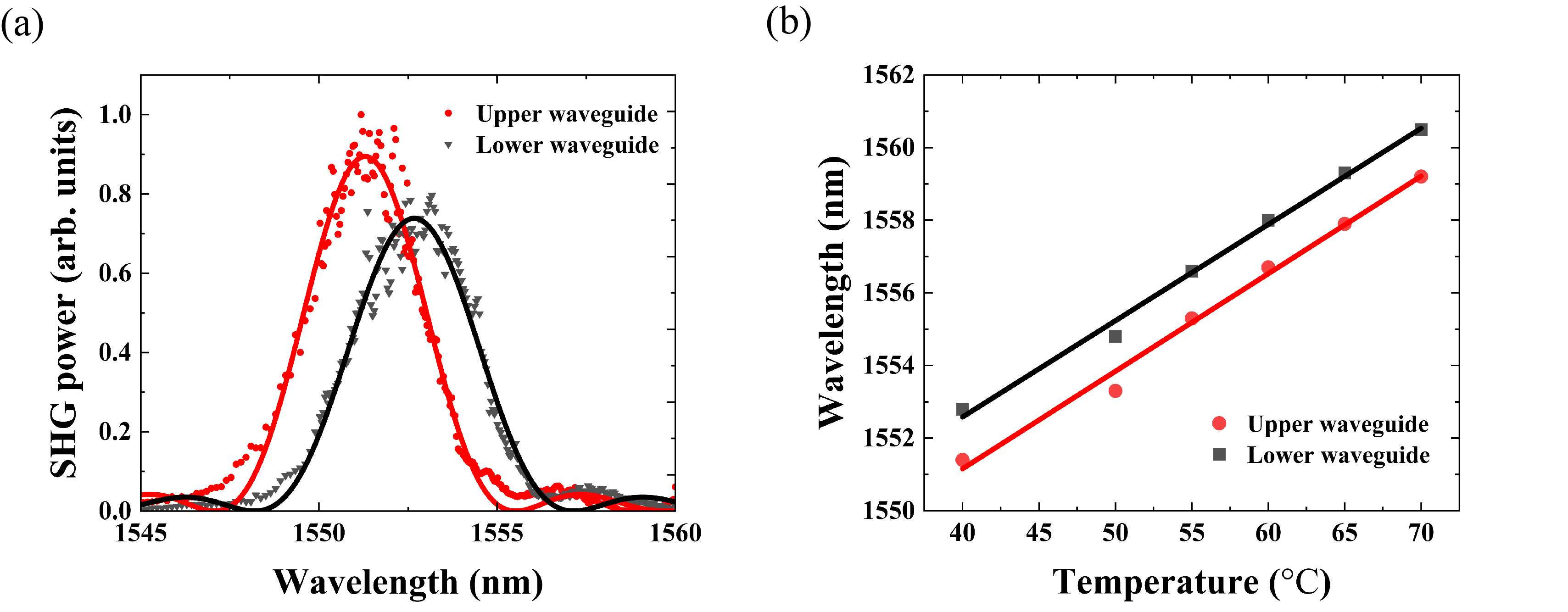}
\caption{The experimental results of second-harmonic generation (SHG) in the dual-channel PPLN waveguides on the TFLN chip:
(a) SHG tuning curve of the dual-channel. The red and gray dots correspond to the measured SHG power for upper and lower waveguides,
respectively. The corresponding fitting curves are plotted in red and black lines. (b) The temperature-dependent behavior of phase-matching wavelengths, with the black and red curves representing the lower waveguide and the upper waveguide, represently.
}
\label{fig:SHG_result}
\end{figure}
For on-chip integrated EPSs, the uniformity between the dual channel PPLN waveguides is critical as it directly governs the EPS performance. To determine the phase-matching wavelengths and verify the consistency of nonlinear properties between the upper and lower waveguides, SHG characterizations are  performed. Continuous-wave fundamental light in the telecom band is coupled into the output port (upper waveguide) of the TFLN chip via an SMF-28 lensed fiber with focal spot size of 2.5 \(\mu\)m. A fiber polarization controller is used to adjust the input light polarization for TE or TM mode coupling. When horizontally polarized, the input light couples into the fundamental TE mode in the upper waveguide and excites the SHG process, generating frequency-doubled light in the 780 nm band. Conversely, vertically polarized light couples into the fundamental TM mode in the upper waveguide, undergoes polarization rotation to the horizontal state, couples into the lower waveguide, and subsequently triggers SHG, producing analogous 780 nm-band frequency-doubled light.  The temperature-dependent phase-matching wavelength shifts are quantified as \(\frac{\Delta \lambda}{\Delta T}\) =
0.269 nm/\(^{\circ}\)C (upper waveguide) and 0.261 nm/\(^{\circ}\)C (lower waveguide) by adjust the TFLN chip to different temperatures via thermo electric cooler (TEC), as shown in Fig.~\ref{fig:SHG_result}(b). Considering the limited wavelength tuning range (\(\sim\)0.8 nm) of the DFB laser chip, we adjust the working wavelength of the TFLN chip by temperature tuning, thereby adapting to the output wavelength of the DFB chip. The operating temperature of the TFLN chip is stabilized at 67 °C using a TEC. By sweeping the fundamental light wavelength, the tuning curves of phase-matching conditions are plotted in Fig.~\ref{fig:SHG_result}(a). The observed SHG efficiency differences of \(\sim\) 1 dB between the dual channel PPLN waveguides mainly stems from the propagation loss difference associated with the TE and TM modes during transmission through the PRC. The dual channel PPLN waveguides exhibit full width at half maxima (FWHM) SHG bandwidths of
\(\Delta\lambda_{\textrm{upper}}\) \(\approx\) 3.85 nm
and \(\Delta\lambda_{\textrm{lower}}\) \(\approx\) 4 nm
, with peak wavelengths at 1558.50 nm and 1559.67 nm, respectively, corresponding to a 1.17 nm mismatch. Such discrepancies are likely due to thickness variations on the TFLN wafer and nonuniformities in the poling and etching processes. The upper waveguide demonstrated a normalized SHG conversion efficiency of 3300\%/(W·c\(\mathrm{m^{2}}\)) with an on-chip fundamental power of 1 mW, reaching 71.8\% of the theoretical efficiency of 4590\%/(W·c\(\mathrm{m^{2}}\)). 
 Despite the tuning curve offsets, their broad FWHM ensures significant spectral overlap, enabling matched SPDC processes under identical pump wavelengths. Concurrently, the working wavelength of the DFB chip is adjusted to 779.5 nm by regulating its working temperature to 21°C via the TEC. In this configuration, a drive current of 40 mA is applied to the DFB laser chip, resulting in 7.5 mW emitted pump power. The maximum coupling efficiency measured through a single straight waveguide is \(\sim\) 10\% primarily limited by the mode size mismatch between the TFLN waveguide mode and the output mode of the DFB laser, yielding a maximum on-chip pump power of 0.75 mW.

\section*{Supplementary Note 4: Broadband photon-pair generation}

\begin{figure}[ht]
\centering
\includegraphics[width=0.6\linewidth]{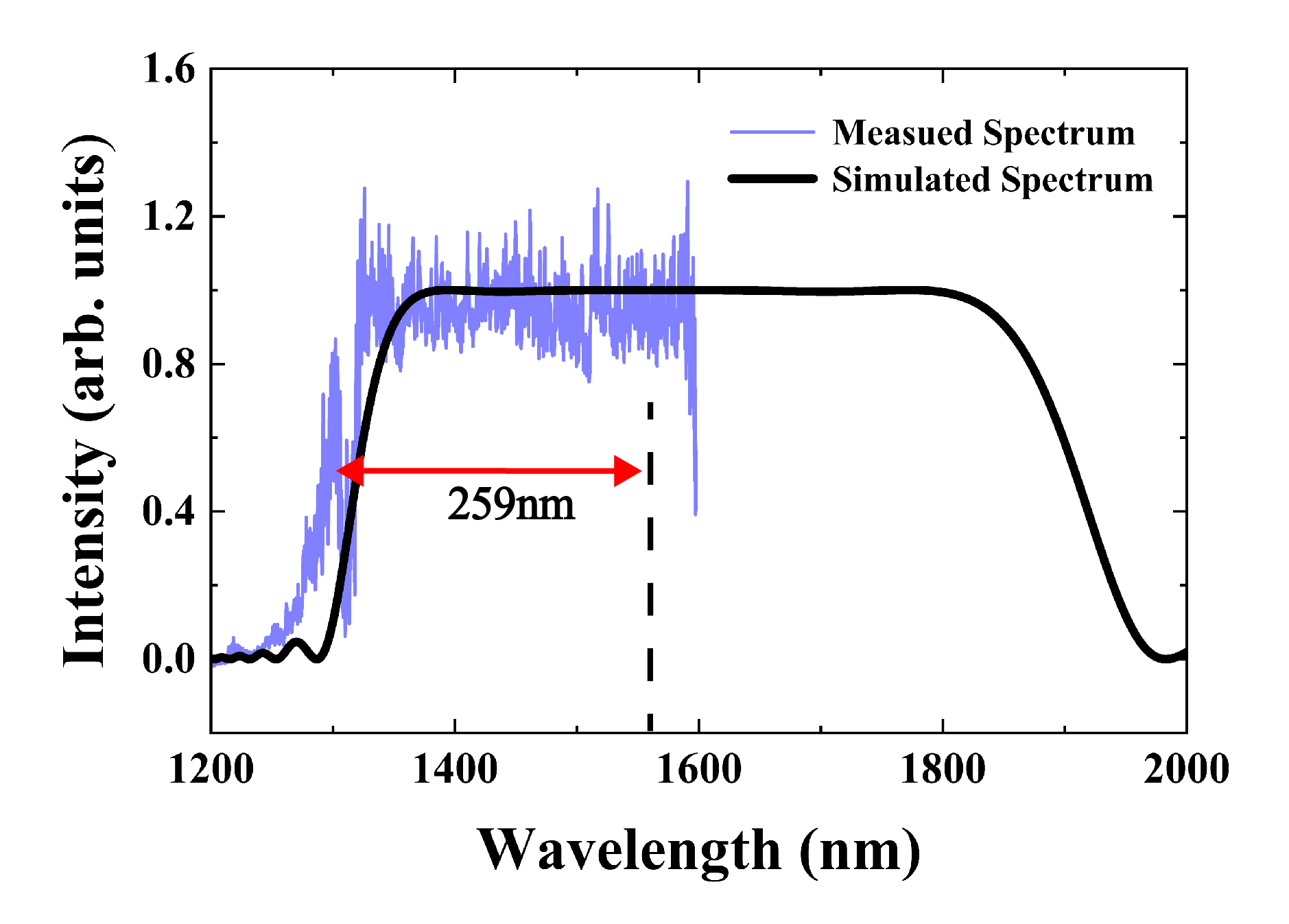}
\caption{Simulated spectrum of SPDC in PPLN waveguide and measured spectrum of SPDC in upper PPLN waveguide. The measured spectrum exhibits a bandwidth of 259 nm, ranging from 1300 nm to 1559 nm.
}
\label{fig:spectrum0}
\end{figure}
The complete SPDC spectrum is measured for horizontally
 polarized photon pairs generated in the upper
waveguide, which are unaffected by the PRC device. The
simulation data demonstrate that the optimized GVD in
the quasi phase matching waveguides enables the generation
 of SPDC photons with broadband spectra, achieving
FWHM bandwidths of 589 nm. In contrast, the experimental
 results reveal that the signal spectral bandwidth
is 259 nm for horizontally polarized light in Fig. 7, indicating
 a SPDC bandwidth of \(\sim\) 600 nm according to
the principle of energy conservation. These measured results
 conffrm that the GVD in the waveguides has been
successfully engineered to facilitate broadband SPDC
emission, thereby enabling the generation of broadband
time–energy entangled photons.

\section*{Supplementary Note 5: PRC measurement}

\begin{figure}[ht]
\centering
\includegraphics[width=0.7\linewidth]{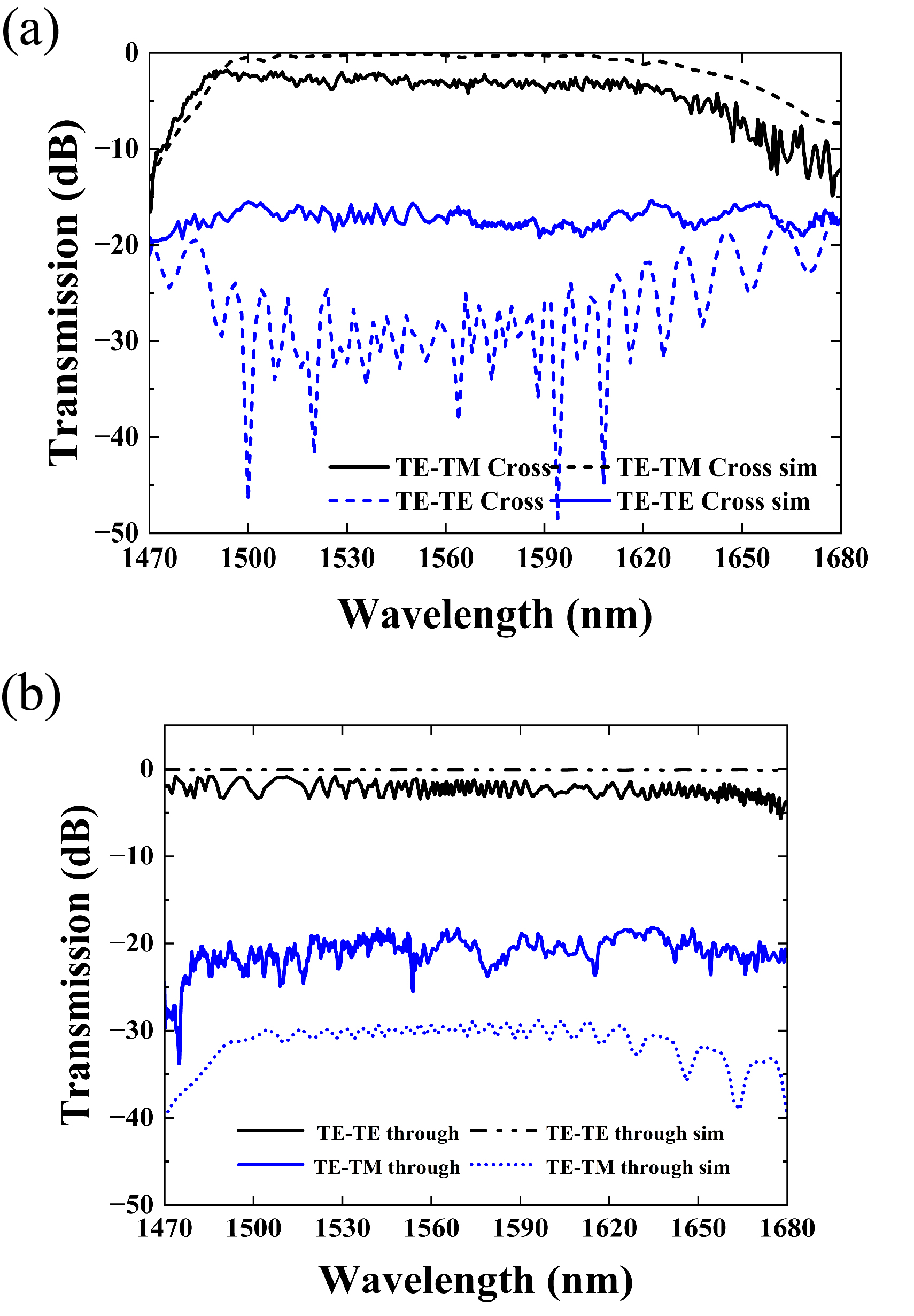}
\caption{Performance of the on chip PRC: (a) Simulated and measured transmissions of horizontal light in the lower waveguide couples into the upper waveguide. (b) Simulated and measured transmissions of horizontal light propagates in the upper waveguide.}
\label{fig:prc}
\end{figure}

To quantitatively characterize the integrated TFLN chip, comprehensive evaluation of PRC is conducted. 
The properties of the PRC is critical for the performance of the polarization-EPS, including bandwidth, brightness and fidelity. To characterize the PRC, the transmission spectrum, insertion losses, and polarization extinction ratios across waveguides are measured. Continuous-wave light from a broadband tunable laser operating within the range of 1470 nm to 1680 nm is coupled into an independent PRC chip via an SMF-28 lensed fiber with a focal spot size of 2.5 \(\mu\)m. A fiber polarization controller is employed to align the input light polarization to the horizontal orientation. At the output port, an objective lens collects the transmitted light, while a polarizer isolates the vertical and horizontal polarization components. Then the output powers from both the upper and lower waveguides are measured individually using a power meter while tuning the wavelength across the 1470–1680 nm range, as shown in Fig.~\ref{fig:prc}(a) and Fig.~\ref{fig:prc}(b). 

After accounting for the fiber-to-chip coupling loss(about -5.5dB on average), the on-chip transmission characteristics exhibit distinct behavior for horizontal and vertical polarizations. Specifically, the PRC exhibits an insertion loss of approximately 2.5 dB for horizontal polarization and 3.3 dB for vertical polarization. The corresponding polarization extinction ratios have been measured as 19.3 dB for horizontally polarized light and 15.5 dB for vertically polarized light, respectively.
The discrepancies observed between simulation and experimental results arise primarily due to two factors: firstly, the fundamental TE and TM modes in the waveguide exhibit cross-polarization components (TM modes contain TE components and vice versa), and secondly, the coupled light is not strictly horizontally polarized.  These findings indicate that entangled photon pairs generated via SPDC in the dual waveguides experience a differential loss profile as they traverse the PRC. This results in amplitude non-uniformity between their respective quantum states \(\mathrm{|H_{s} H_{i}\rangle}\) and \(\mathrm{|V_{s} V_{i}\rangle}\). Furthermore, the imperfect polarization extinction ratios give rise to additional quantum states, namely \(\mathrm{|V_{s} H_{i}\rangle}\) and \(\mathrm{|H_{s} V_{i}\rangle}\), which further complicate the overall system behavior.
The polarization conversion spectrum from the lower waveguide (T\(\rm{E_0}\)) to the upper waveguide (T\(\rm{M_0}\)) demonstrates a broad bandwidth, with a full width at half maximum (FWHM) of 160 nm, which is determined by the initial (\(w_{4}\)) and final (\(w_{5}\)) widths of the lower waveguide. The geometry-dependent bandwidth is critical to determine the operating bandwidth of the EPS. Importantly, both the loss characteristics and bandwidth performance of the PRC can be systematically optimized by precisely tuning the waveguide geometry parameters.

\end{document}